\documentclass[prl,twocolumn,superscriptaddress,amsmath,amssymb]{revtex4}
\usepackage{graphicx}

\renewcommand{\vec}[1]{{\bf #1}}
\newcommand{\erf}{\mathop{\rm erf}}
\newcommand{\erfc}{\mathop{\rm erfc}}
\newcommand{\rmi}{{\rm i}}
\begin{document}
\title{Intermediate wave-function statistics}
\author{G.\ Berkolaiko}
\affiliation{Department of Mathematics, University of Strathclyde, 
Glasgow G1 1XH, UK.}
\author{J.\ P.\ Keating}%
\author{B.\ Winn}
\email{b.winn@bristol.ac.uk}
\affiliation{School of Mathematics, University of Bristol, 
Bristol BS8 1TW, UK.}%
\date{August 5, 2003}
\begin{abstract}
We calculate statistical properties of the eigenfunctions of two quantum
systems that exhibit intermediate spectral statistics: star graphs and
\v{S}eba billiards. First, we show that these eigen\-functions are not
quantum ergodic, and calculate the corresponding limit distribution.
Second, we find that they can be strongly scarred, in the case of star graphs
by short (unstable) periodic orbits and in the case of \v{S}eba billiards 
by certain families of orbits. We construct sequences of states
which have such a limit. Our results are illustrated by numerical
computations.
\end{abstract}
\pacs{Valid PACS appear here}
\maketitle

It has been conjectured that the quantum spectral statistics of systems that
are chaotic in the semi-classical limit are generically those of Random Matrix 
Theory
\cite{boh:ccq}. The behaviour of the eigenfunctions of such systems is
described by the semi-classical eigenfunction hypothesis 
\cite{berry:77, vor:sce}, which implies that they 
equidistribute over the appropriate energy shell. This is in agreement with a 
theorem of Schnirelman \cite{sch:epe} which implies equidistribution of
almost all eigenstates on scales independent of $\hbar$, assuming only 
classical ergodicity. Such behaviour is termed {\em quantum ergodicity}.
This theorem still permits the possibility of a small number of states which
do not equidistribute. 

It has been suggested that some of these exceptional states may be
``scarred'' by short classical periodic orbits \cite{hel:bse}.
Further investigations 
\cite{bog:swf, ber:qsc, kea:obs, aga:scs, kap:sqc,sch:scn} 
have distinguished between weak and strong scarring.
Weak scarring relates to states averaged over energy windows that
contain a semi-classically increasing number of levels,
whereas strong scarring means that sequences
of states can be constructed whose  limit is wholly or in part supported by 
one-or-more periodic orbits. So far the only systems known
rigorously to support strong scarring are the cat maps \cite{fau:sef},
which have non-generic spectral statistics \cite{kea:tcm}. 

For systems that are classically integrable, it is expected that the 
quantum spectral statistics are Poissonian, i.e.\ those of independent random
numbers \cite{ber:lcr}. The corresponding eigenfunctions semiclassically 
equi\-distribute on tori in phase space \cite{ber:scm}.

Recently, classes of systems which exhibit spectral statistics that are
intermediate between Random Matrix and Poissonian have been 
discovered \cite{shk:ssd, bog:ss, bog:srp}. Two representative families of
examples are \v{S}eba billiards \cite{seb:wcs} and star graphs \cite{ber:tps}.
It was shown in \cite{ber:sgs} that these two systems have the same
(intermediate) spectral statistics.
We study the eigenfunction statistics of such systems. Specifically, given that
these systems are not classically ergodic, we are interested in whether
the eigenfunctions are quantum ergodic, and whether they show strong scarring
(that they exhibit weak scarring may be shown using the methods of
\cite{bog:swf}).

Star graphs are quantum graphs \cite{kot:pot} which have one central vertex,
and $b$ outlying vertices each connected only to the central vertex
\cite{ber:tps}. 
For such graphs, the limit $b\to\infty$ is
analogous to the semi-classical limit. 
To investigate the possibility of quantum ergodicity
in this limit, we consider a graph with $b=\alpha v$ bonds, where $v\gg 1$,
$\alpha>1$, and introduce the observable $B$ defined by
\begin{equation*}
  B=\left\{ 
  \begin{array}{ll}
     1 & \mbox{on bonds indexed $1,\ldots,v$} \\
    0 & \mbox{on bonds indexed $v+1,\ldots,b$.}
  \end{array}
\right.
\end{equation*}
Thus $B$ picks out a fraction $\alpha^{-1}$ of the bonds.
Let $\psi_n$ denote the wave-function associated with the
$n^{\rm th}$ eigenstate. 
We calculate the probability distribution, $P(R)$, for $n$ chosen at
random, that $\langle\psi_n|B|\psi_n\rangle$ is less
than $R$, subject to some mild restrictions on the bond 
lengths. A system which exhibits quantum ergodicity would have
\begin{equation*}
  P(R)=\left\{
 \begin{array}{ll}
     0, & \qquad 0\leq R<\alpha^{-1} \\
     1, & \qquad \alpha^{-1}\leq R\leq 1.
  \end{array}
\right.
\end{equation*}
Our result (see equation (\ref{eq:6}) and figure \ref{fig:2} below) differs 
from this, proving that star graphs are not quantum ergodic.
In fact we are able to say more: for a fixed (finite) number of bonds,
we explicitly find eigenstates that are strongly scarred along closed 
(unstable)
orbits of the graph with period 2. This is the first class of examples showing
generic (in this case intermediate) behaviour in which strong scarring has 
been rigorously demonstrated.

The term {\em \v{S}eba billiard} refers to any integrable quantum system that
has been perturbed by the addition of a point singularity. We consider
the specific example of a billiard on a torus. By exploiting the connection
between \v{S}eba billiards and star graphs \cite{ber:sgs} we argue that
\v{S}eba billiards are also not quantum ergodic and find states that
appear to show behaviour analogous to strong scarring, in this case by families
of orbits.

We begin by describing how to calculate the probability distribution $P(R)$.

Eigenenergies of a star graph with $b$ bonds are given by
$E_n=k_n^2$, where $k_n$ is the $n^{\rm th}$ solution of $Z(k)=0$ with
\begin{equation}
  \label{eq:spec_det}
  Z(k)=\sum_{j=1}^{b} \tan kL_j,
\end{equation}
the individual bond lengths being denoted by $L_1,\ldots,L_{b}$.
The component of the $n^{\rm th}$ wave-function on the $i^{\rm th}$ 
bond of the graph is $\psi_{n,i}(x) = A_i(k_n) \cos k_n(x-L_i)$, where
\begin{equation}
  \label{eq:estate}
  A_i(k_n) = \frac{\sqrt{2}}{\cos k_nL_i(\sum_{j} L_j
    \sec^2 k_nL_j)^{1/2}},
\end{equation}
the sum being taken over all bonds.
Then
\begin{equation}
  \langle\psi_n|B|\psi_n\rangle=\frac{\sum_{i=1}^v L_i\sec^2k_nL_i}
{\sum_{j=1}^{b} L_j\sec^2k_nL_j}+O(k_n^{-1}).
\label{eq:7}
\end{equation}
To calculate the distribution of values
taken by this quantity
we average over a large number of states, making the error term in (\ref{eq:7})
negligible. We choose incommensurate bond lengths from an interval
$[\bar{L},\bar{L}+\Delta L]$ that shrinks in such a way that
$v\Delta L\to 0$ as $v\to\infty$. Thus we can replace
$L_i$ by $\bar{L}$ wherever it does not multiply $k_n$.

To evaluate a function $f(k)$ at the zeros of $Z(k)$ we integrate against
the density of states, so
\begin{equation*}
  \frac{1}{N}\sum_{n=1}^{N}f(k_n) =\frac{1}{N}\int_0^{k_N} f(k)Z'(k)
\delta[Z(k)] dk
\end{equation*}
where $\delta$ denotes the Dirac delta function. Writing the delta function
in Fourier representation, $\delta(x)=(2\pi)^{-1}\int_{-\infty}^{\infty}
e^{\rmi\zeta x}d\zeta$, and taking the limit $N\to\infty$,
{\setlength\arraycolsep{1pt}
\begin{eqnarray} \label{eq:2}
  \lim_{N\to\infty}&&\frac{1}{N}\sum_{n=1}^{N}f(k_n)=\\
&&\frac{1}{2\pi\bar{d}}\lim_{K\to\infty}\frac{1}{K}\!\int_{0}^K
\int_{-\infty}^{\infty} f(k)Z'(k)\exp[\rmi\zeta Z(k)]d\zeta dk,
\nonumber
\end{eqnarray}}
$\!\!$writing $K=k_N$ and using $k_N\approx N/\bar{d}$, where 
$\bar{d}=b\bar{L}/\pi$ is the mean density of states.
We apply (\ref{eq:2}) with $f(k)=\exp(\rmi\beta X_{\eta}(k))$ where 
\begin{equation}
X_{\eta}(k)=\frac{1}{v^2}\sum_{j=v+1}^{b}
\sec^2 kL_j-\frac{\eta}{v^2}\sum_{i=1}^{v}\sec^2 kL_i
\end{equation}
for $\beta, \eta$ constants. This is related to the distribution 
of $\langle\psi_n|B|\psi_n\rangle$ by the fact that
\begin{equation*}
  {\mathbb P}(X_{\eta}(k_n)>0) = {\mathbb P}(\langle\psi_n|B|\psi_n\rangle<R) 
\end{equation*}
when $R$ and $\eta$ are related by $\eta=1/R-1$.

We observe that $k$ only appears in (\ref{eq:2}) multiplied by a
bond length, and as an argument of a $\pi$-periodic function. Since 
the bond lengths are incommensurate, the $k$ integral can be
re-written as a multiple integral over the $b$ variables $x_j=kL_j$.
A similar argument was used in \cite{bar:lsd, kea:vde}. The integrand then
factorises, so that
\begin{eqnarray}
 \lim_{N\to\infty}\frac{1}{N}\sum_{n=1}^{N}f(k_n)&=&
\frac{1}{2\alpha v}\int_{-\infty}^{\infty}I_1 I_2^{v-1} I_3^{\alpha v-v}
\nonumber\\
&&+(\alpha-1)I_4 I_2^v I_3^{\alpha v -v -1} d\zeta
\label{eq:4}
\end{eqnarray}
where
{\setlength\arraycolsep{2pt}
\begin{eqnarray*}
  I_1&=& \frac{1}{\pi}\int_0^{\pi}\sec^2 x \exp\left(\frac{\rmi\zeta}{v}
\tan x-\frac{\rmi\beta\eta}{v^2}\sec^2 x\right)d x, \\
 I_2&=& \frac{1}{\pi}\int_0^{\pi}\exp\left(\frac{\rmi\zeta}{v}
\tan x-\frac{\rmi\beta\eta}{v^2}\sec^2 x\right)d x,
\end{eqnarray*}}
$\!\!I_3$ is obtained by replacing $\beta$ with $-\beta/\eta$ in $I_2$, and
$I_4$ by making the same substitution in $I_1$.
Techniques to analyse the asymptotics of these integrals were discussed 
in \cite{kea:vde}. Using them we find that
\begin{equation*}
  I_1\sim\frac{v}{\sqrt{\pi\rmi\beta\eta}}\exp\left(\frac{\rmi\zeta^2}
{4\beta\eta}\right),
\end{equation*}
and
\begin{equation*}
   I_2^{v}\sim\exp\left[-\frac{2}{\sqrt{\pi}}\sqrt{\rmi\beta\eta}\exp\left(
\frac{\rmi\zeta^2}{4\beta\eta}\right)-
\zeta\erf\left(\frac{\zeta}{2\sqrt{\rmi\beta\eta}}\right)
\right],
\end{equation*}
as $v\to\infty$. Substituting the above into (\ref{eq:4}) and denoting the 
result $e(\beta)$, we arrive at
\begin{equation*}
e(\beta)=\frac{1}{2\alpha}\int_{-\infty}^{\infty}\frac{1}{\sqrt{\beta}}
T\!\left(
\frac{\zeta}{\sqrt{\beta}}\right)\exp\left[-\sqrt{\beta}\tau\!
\left(\frac{\zeta}{\sqrt{\beta}}\right)\right]d\zeta,
\end{equation*}
where
\begin{equation*}
  T(\xi)=\frac{1}{\sqrt{\rmi\pi\eta}}\exp\left(
\frac{\rmi\xi^2}{4\eta}\right)+\frac{(\alpha-1)}{\sqrt{-\rmi\pi}}\exp
\left(-\frac{\rmi\xi^2}{4}\right)
\end{equation*}
and
\begin{eqnarray*}
  \tau(\xi)&=& \frac{2}{\sqrt{\pi}}\sqrt{\rmi\eta}\exp\left(
\frac{\rmi\xi^2}{4\eta}\right)+\xi\erf\left(
\frac{\xi}{2\sqrt{\rmi\eta}}\right) \\
&+&\frac{2(\alpha-1)}{\sqrt{\rmi\pi}}\exp\left(
-\frac{\rmi\xi^2}{4}\right)+\xi(\alpha-1)\erf\left(
\frac{e^{\rmi\pi/4}\xi}{2}\right).
\end{eqnarray*}
The Fourier transform of $e(\beta)$ is the probability density function
of $X_{\eta}(k_n)$
where the index of the state, $n$, is chosen at random. The probability 
distribution for $\langle\psi_n|B|\psi_n\rangle$ to be less than $R$
is then given by  
\begin{equation}
  P(R)=\left.\frac{1}{2\pi}\int_{0}^{\infty}\int_{-\infty}^{\infty} e(\beta)
e^{-\rmi\beta\sigma} d\beta d\sigma\right|_{\eta=1/R-1}
\label{eq:8}
\end{equation}

The Fourier transform of $e(\beta)$ is
\begin{equation*}
  \frac{-1}{2\pi\alpha}\Re\int_{-\infty}^{\infty}
T(\xi)\frac{
\sqrt{\pi}\tau(\xi)}{2(\rmi\sigma)^{3/2}}
w\left(\frac{-\tau(\xi)}{2\sqrt{-\rmi\sigma}}
\right)d\xi,
\end{equation*}
having made the substitution $\xi=\zeta/\sqrt{\beta}$ and using the notation
$w(z)=e^{-z^2}\erfc(-\rmi z)$.
Performing the $\sigma$-integral in (\ref{eq:8}) gives, finally,
\begin{equation}
  P(R)=\frac{1}{2}-\frac{1}{\pi\alpha}\Im\int_{-\infty}^{\infty} T(\xi)
\log(\tau(\xi)) d\xi
\label{eq:6}
\end{equation}
with $\eta=1/R-1$, for $0<R<1$.

\begin{figure}
\includegraphics[angle=0,width=8.0cm,height=6cm]{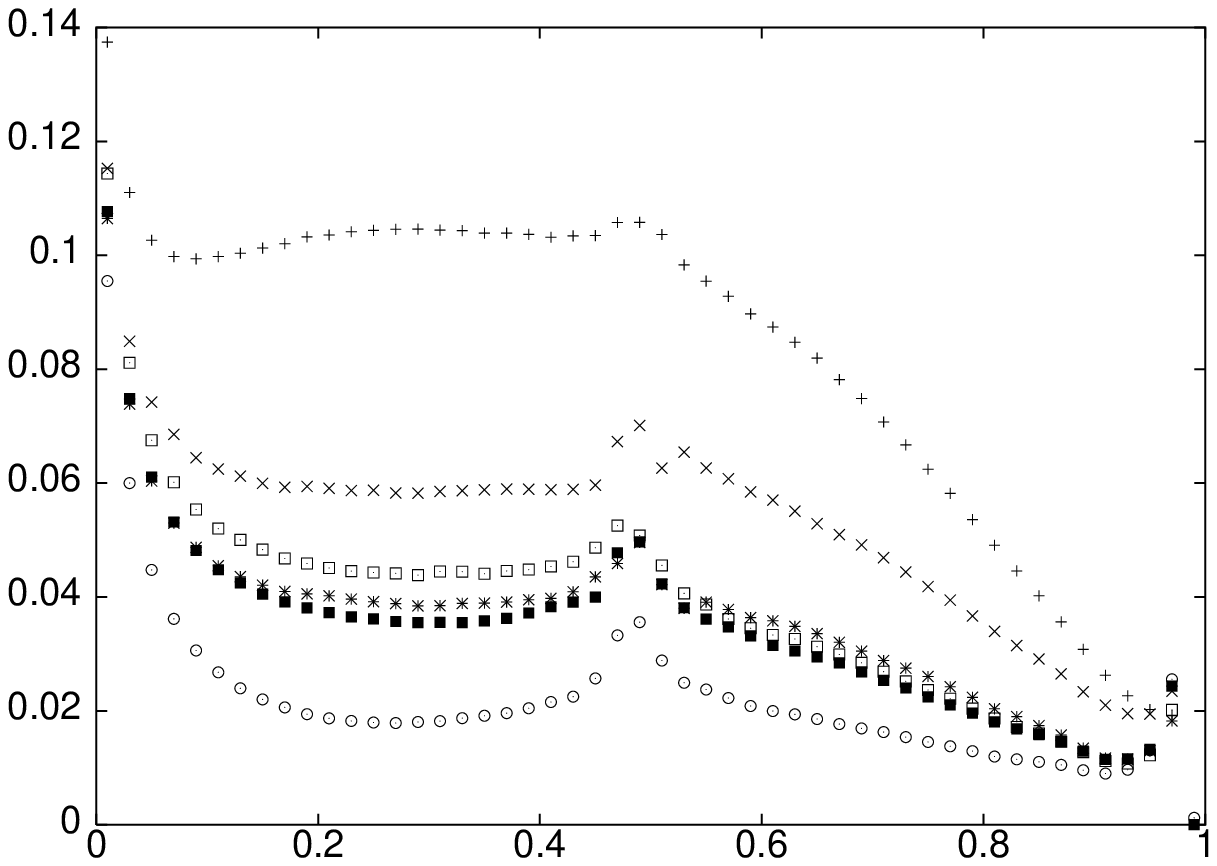}
\caption{Difference between $P(R)$ and numerics when 
$b=15 (+),$ $30 (\times), 
45 (+\mspace{-14mu}\times), 
60 (\boxdot), 75 (\blacksquare), 90 (\odot)$.}
\label{fig:99}
\end{figure}

The results of numerical computations which support this calculation are 
shown in figures  \ref{fig:2} and \ref{fig:99}.

\begin{figure}[t]
\includegraphics[angle=0,width=8.0cm,height=6cm]{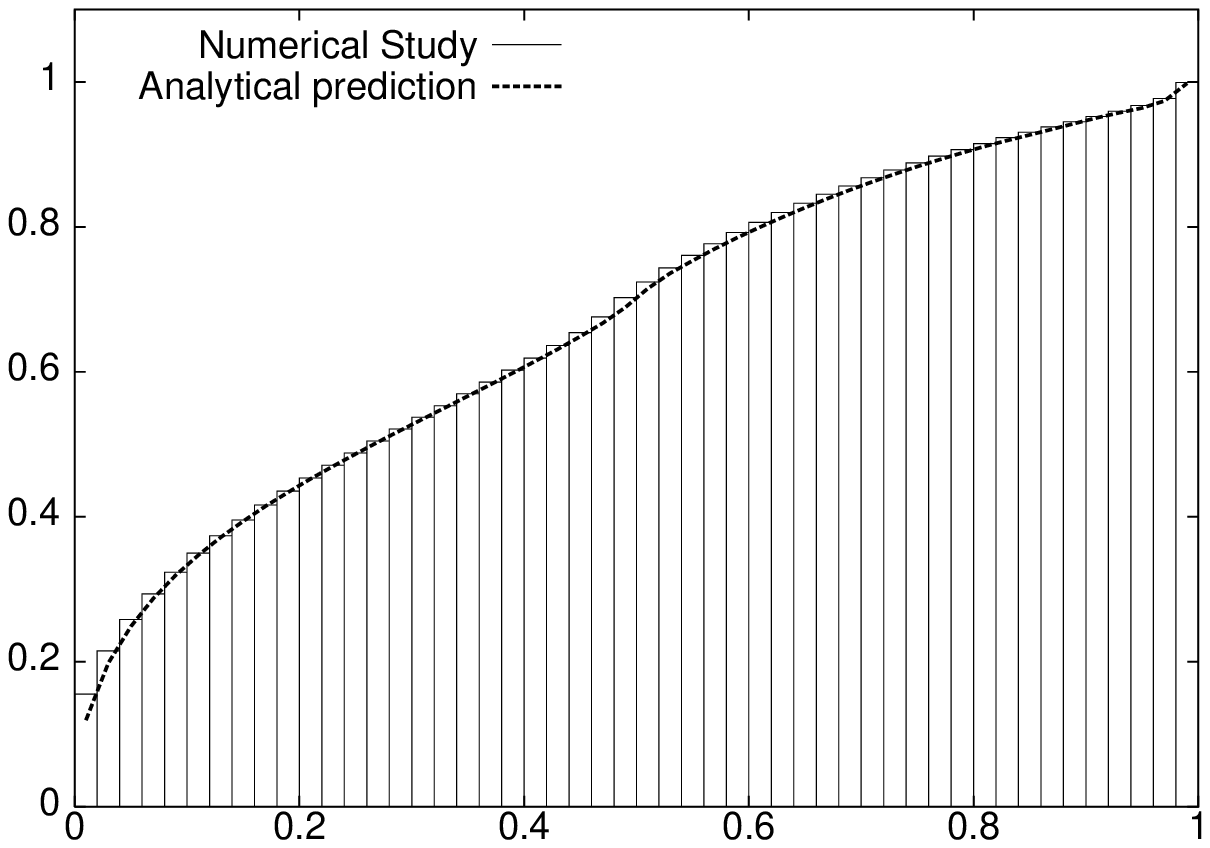}
\caption{Comparing $P(R)$, as given by (\ref{eq:6}), to a direct numerical 
computation for a star graph with 90 bonds when $\alpha=3$.}
\label{fig:2}
\end{figure}

We now turn to constructing sequences of eigenstates on star graphs, when 
$b$ is
fixed, that are strongly scarred by certain short periodic orbits. (Note that
on such graphs all orbits are unstable.)  Our construction exploits the
properties of the spectral determinant (\ref{eq:spec_det}).
The spectral determinant has poles at the points 
\begin{equation*}
  P = \bigcup_{i=1}^{b} P_i \equiv \bigcup_{i=1}^{b} \left\{
  \frac{\pi/2 + \pi n}{L_i} : n\in{\mathbb Z} \right\}.
\end{equation*}  
Since the derivative of $Z(k)$ is everywhere greater than zero, there
is exactly one root of $Z(k)=0$ between every two consecutive poles.

Given a small $\epsilon>0$, which will control the quality of the
scarred eigenstate, we can find a pole $p_1$ in the set $P_1$
satisfying the following properties: (a) there is a pole $p_2$ from
$P_2$ within a distance $\epsilon$ of $p_1$ and (b) $p_1$ is
approximately equidistant from the two nearest poles from $P_i$, for
each $i>2$.  Due to the ergodic properties of the sequence $P$
(assuming that the bond lengths $L_i$ are incommensurate), 
the above situation occurs
with non-zero frequency along the $k$-axis.

Denote the root squeezed between $p_1$ and $p_2$ by $k'$.  Then
$\cos k'L_i$ is of the order of $\epsilon$ when $i=1,2$ and 
is of order $1$
otherwise.  Going back to the eigenstate formula (\ref{eq:estate}), we
see that 
\begin{equation*}
  \frac{A_{1,2}(k')}{A_i(k')} =O(\epsilon^{-1}) \qquad \mbox{ for } i>2,
\end{equation*}
that is, the amplitude of the $k'$-eigenstate on the bonds 1 and 2 is
$\epsilon^{-1}$ times stronger than on any other bond.  By selecting
suitably small $\epsilon$ one can find eigenstates localized on any two given
bonds to any precision.  Understandably, higher precision leads to
a smaller frequency of the scarred eigenstates.  In fact, the frequency is
proportional to $\epsilon$.

Since $Z(k')=0$ it follows that $A_1(k') \approx A_2(k')$ which 
provides an
explanation for the visible singularity at $R=1/2$ in the difference
between $P(R)$ for finite $b$ and its limiting form (see
figure \ref{fig:99}).  This singularity corresponds to the
eigenstates localized on bonds $e$ and $e'$ such that $e$ is picked
out by the observable $B$ and $e'$ is not.

The above construction can be generalized to produce eigenstates
localised on any number $j\geq2$ of bonds.  However, once $j>2$, the
amplitudes on the $j$ bonds are generally not equal, which explains
the lack of singularities at rational fractions other than $1/2$.
Finally, the singularities at $R=0$ and $1$ correspond to the cases
when the eigenstates are localized fully outside ($R=0$) or inside ($R=1$) the
$v$ bonds picked out by $B$.  

The preceding calculations can be made rigorous. We defer the details
to \cite{kea:nqe}.

In \cite{kea:vde} it was suggested that the squares of the coefficients, 
$c_i^2$, of the eigenfunctions of \v{S}eba billiards expressed in the basis of 
states of the unperturbed billiard
\begin{equation}
  |\psi\rangle=\sum_i c_i |\psi^{(0)}_i\rangle
\end{equation}
are distributed in the same way as
the square of the maximum norm on a single bond of a star graph in
the limit as $v\to\infty$. This conjecture was supported by numerical
evidence.
We extend this analogy to interpret the above results
in terms of the \v{S}eba billiard. 
Since the quantity in (\ref{eq:7})
is similar to a sum of norms of eigenfunctions on a fraction of bonds,
we conjecture that the sum of the squares of a fraction $\alpha^{-1}$
of the coefficients has probability distribution $P(R)$. 
To elucidate this idea, consider preparing a \v{S}eba-type
system in a randomly-chosen eigenstate. 
The perturbation is then removed instantaneously, and a measurement of 
the energy is made. 
What is the distribution (with respect to the choice of 
initial state) of the probability that the measured energy is 
one of a given fraction $\alpha^{-1}$ of the energy levels of the unperturbed 
system? The answer is the distribution function in (\ref{eq:6}).
If the eigenfunctions of the billiard were asymptotically 
equidistributed then this probability distribution would be a 
unit step function at $R=1/\alpha$. 

Energy levels of a \v{S}eba billiard interlace with energy levels of
the original unperturbed system in much the same way that momenta
of star graphs interlace with poles of the function $Z(k)$.
We consider a Neumann billiard in a rectangle with aspect ratio $\gamma^{1/2}$,
perturbed by a point singularity at the origin. 
Eigenstates of this system can be expanded as
\begin{equation}
 |\psi_n(\vec{x})\rangle =A_n\sum_{i,j}\frac{|\psi_{i,j}^{(0)}(\vec{x})\rangle}
{E^{(0)}_{i,j}-E_n}
\end{equation}
where $A_n$ is a normalisation constant, the energy
levels of the Neumann billiard are
$E_{i,j}^{(0)}=4\pi^2\gamma^{-1/2}(i^2+\gamma j^2)$, 
and $|\psi_{i,j}^{(0)}\rangle$ are
the corresponding eigenfunctions. It is well known that these 
unperturbed eigenfunctions
are localised in momentum space. We therefore expect to find states
of the \v{S}eba billiard that exhibit structures analogous to scars 
in momentum space when their
energy is between two closely spaced levels of the unperturbed
billiard. In fact such states will scar in two directions in momentum space,
corresponding to the two unperturbed eigenstates closest in energy to the 
state in question.
These scars are supported by families of orbits corresponding to tori in
the unperturbed system. Note however that torus quantisation itself does
not apply. It is in this sense that the structures are analogous to scars.

\begin{figure}[t!]
  \def\fscale{0.64}
  \includegraphics[scale=\fscale]{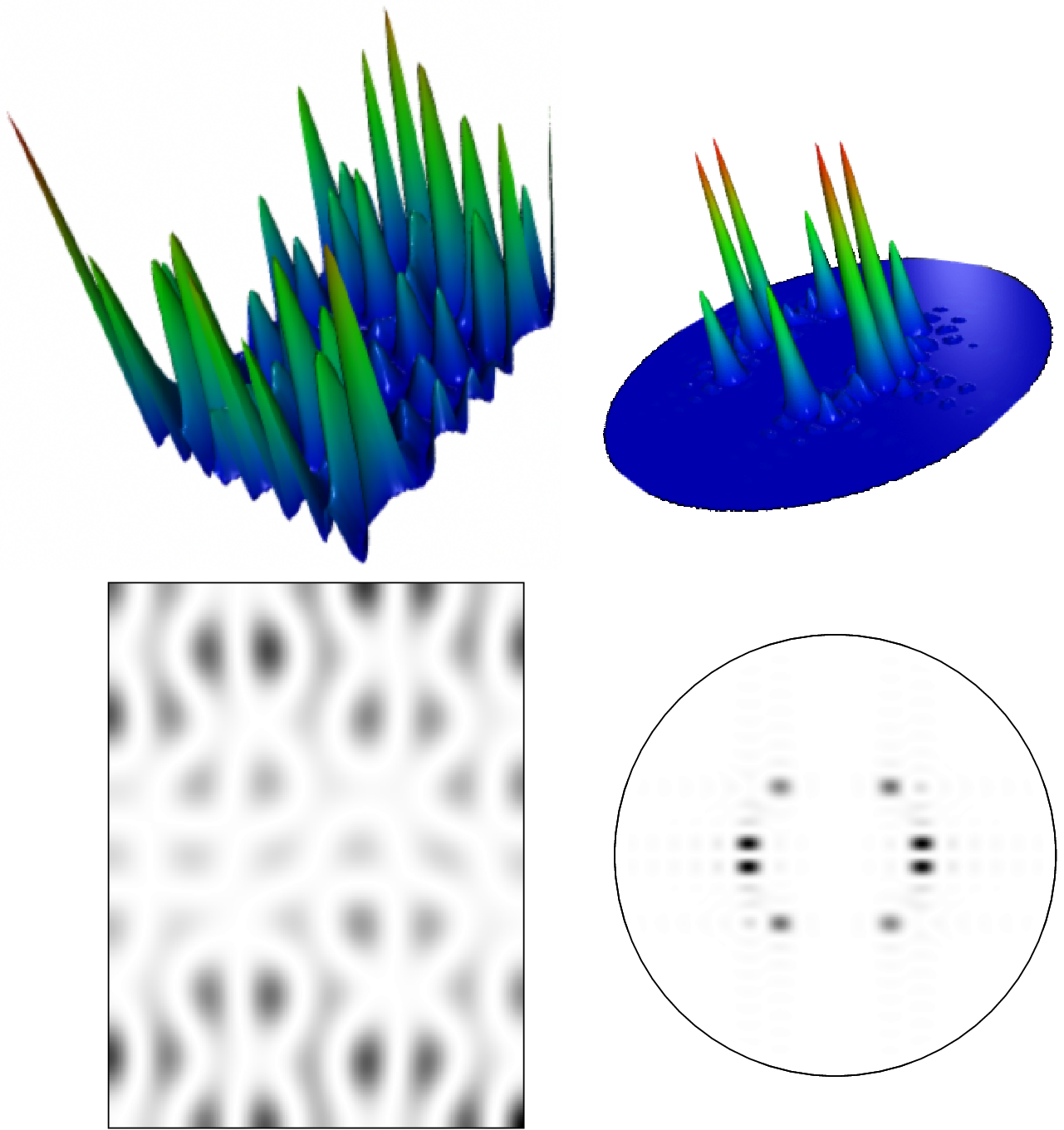} 
  \caption{The wave-density of the $55^{\rm th}$ eigenfunction of 
the \v{S}eba billiard, in position (left) and momentum (right) space.
Intensity plots are shown below the three-dimensional plots;
greater probability density is encoded as darker points.
In this example $\gamma=(\sqrt{5}+1)/2$.}
  \label{fig:1}
\end{figure}

Figure \ref{fig:1} shows the $55^{\rm th}$ state of the \v{S}eba billiard 
described above, with the scatterer placed at the origin. 
Although there is no clear localisation evident in position
representation, the momentum representation clearly shows localisation
in two directions.

BW was financially supported by an EPSRC studentship (Award Number 0080052X).

\end{document}